\documentclass[12pt]{article}
\usepackage{graphicx}
\usepackage{graphics}
\title{Fractional part integral representation for derivatives of a function related to 
$\ln \Gamma(x+1)$} 
\author{Mark W. Coffey\\
Department of Physics\\
Colorado School of Mines\\
Golden, CO  80401\\
(Received $\mbox{~~~~~~~~~~~~~~~~~~~~~~~~~~~~~~~2010}$)}
\date{August 14, 2011}
\begin{document}
\maketitle
\baselineskip=25 pt
\begin{abstract}

For $0\neq x>-1$ let 
$$\Delta(x)={{\ln \Gamma(x+1)} \over x}.$$
Recently Adell and Alzer proved the complete monotonicity of $\Delta'$ on $(-1,\infty)$
by giving an integral representation of $(-1)^n \Delta^{(n+1)}(x)$ in terms of the
Hurwitz zeta function $\zeta(s,a)$.  We reprove this integral representation in 
different ways, and then re-express it in terms of fractional part integrals.  Special
cases then have explicit evaluations.  Other relations for $\Delta^{(n+1)}(x)$ are
presented, including its leading asymptotic form as $x \to \infty$.

\end{abstract}
 
\medskip
\baselineskip=15pt
\centerline{\bf Key words and phrases}
\medskip 

\noindent

Gamma function, digamma function, polygamma function, Hurwitz zeta function, Riemann zeta function, fractional part, integral representation

\vfill
\centerline{\bf 2010 AMS codes} 
33B15, 11M35, 11Y60

\baselineskip=25pt
\pagebreak
\medskip
\centerline{\bf Introduction and statement of results}
\medskip

For $0\neq x>-1$ let 
$$\Delta(x)={{\ln \Gamma(x+1)} \over x}, ~~~~~~\Delta(0)=-\gamma, \eqno(1.1)$$
where $\Gamma$ is the Gamma function, $\gamma=-\psi(1)$ is the Euler constant, and
$\psi(x)=\Gamma'/\Gamma$ is the digamma function.  The study of the convexity and
monotonicity of the functions $\Gamma$ and $\Delta$ and of their derivatives is of
interest \cite{grabner,qi,qiguo,vogt}.  
For instance, the paper \cite{grabner} gave an analog of the well known Bohr-Mollerup theorem for the function $\Delta(x)$.  Monotonicity and convexity are very useful properties for developing a variety of inequalities.  Completely monotonic
functions have applications in several branches, including complex analysis, potential
theory, number theory, and probability (e.g., \cite{bochner}).  In \cite{berg}, $-\Delta(x)$ was shown to be a Pick function, with integral representation
$$-\Delta(x)=-{\pi \over 4}+\int_{-\infty}^{-1} \left({1 \over {t-z}}-{t \over {t^2+1}}
\right){{dt} \over {-t}}.$$
I.e., this function is holomorphic in the upper half plane with nonnegative 
imaginary part.

Recently Adell and Alzer \cite{aa} proved the complete monotonicity of $\Delta'$ on $(-1,\infty)$ by demonstrating the following integral representation.
{\newline \bf Proposition 1}. (Adell and Alzer).  For $x>-1$ and $n \geq 0$ an integer
one has
$$(-1)^n \Delta^{(n+1)}(x)=(n+1)!\int_0^1 u^{n+1}\zeta(n+2,xu+1)du, \eqno(1.2)$$
where $\zeta(s,a)$ is the Hurwitz zeta function (e.g., \cite{ivicbk}).  The complete monotonicity of $\Delta'$, the statement $(-1)^n \Delta^{(n+1)}(x) \geq 0$, then follows
from $\zeta(n+2,xu+1) \geq 0$ for $x >-1$.  We reprove the result (1.2) in two other ways, 
and in so doing illustrate properties of the $\zeta$ function.
\newline{\bf Corollary 1}.  We have the following recurrence:
$${{(-1)^n} \over {(n+1)!}}\Delta^{(n+1)}(x)={1 \over x}{{(-1)^{n-1}} \over {n!}}\Delta^{(n)}(x)-{{\zeta(n+1,x+1)} \over {(n+1)x}}.  \eqno(1.3)$$

We then relate cases of (1.2) to fractional part integrals, including the following,
wherein we let $\{x\}=x-[x]$ denote the fractional part of $x$.
{\newline \bf Proposition 2}.  Let $k \geq 1$ be an integer.  Then we have
$$\int_0^1 u^{n+1}\zeta(n+2,ku+1)du ={1 \over k^{n+2}}\left[\int_1^\infty {{\{w\}^{n+1}}
\over w^{n+2}}dw+\sum_{j=1}^{k-1}\int_0^\infty {{(\{x\}+j)^{n+1}} \over {(x+j+1)^{n+2}}}dx
\right].  \eqno(1.4)$$
As a further special case we have
{\newline \bf Corollary 2}.  We have
$$(-1)^n \Delta^{(n+1)}(1)=(n+1)!\int_0^1 y^{n+1}\zeta(n+2,y+1)dy=(n+1)!\int_0^\infty {{\{x\}^{n+1}}\over {(x+1)^{n+2}}}dx$$
$$=(n+1)!\left[1-\gamma-\sum_{j=2}^{k-1} {1 \over j}[\zeta(j)-1]\right],  \eqno(1.5)$$
where $\zeta(s)=\zeta(s,1)$ is the Riemann zeta function \cite{edwards,ivicbk,riemann,titch}.

More generally, we have the following, wherein we put $P_1(x)=\{x\}-1/2$.  Let
$_2F_1$ be the Gauss hypergeometric function \cite{andrews,grad}.
{\newline \bf Proposition 3}.  We have for integers $n \geq 0$
$$\int_0^1 u^{n+1}\zeta(n+2,xu+1)du ={1 \over {2(n+2)}}{1 \over {(x+1)^{n+2}}} ~_2F_1\left(
1,n+2;n+3;{x \over {x+1}}\right)$$
$$+ {1 \over {(n+1)(n+2)}}{1 \over {(x+1)^{n+1}}} ~_2F_1\left(
1,n+1;n+3;{x \over {x+1}}\right)
-\int_0^\infty {1 \over {(t+1)}}{{P_1(t)} \over {(t+x+1)^{n+2}}}dt.  \eqno(1.6)$$
From this Proposition we may then determine the following asymptotic form:
{\newline \bf Corollary 3}.  We have
$$\Delta^{(n+1)}(x) \sim (-1)^n {{n!} \over {(x+1)^{n+1}}}, ~~~~~~x \to  \infty, \eqno(1.7)$$
in agreement with Corollary 1.2 of \cite{aa}.  In fact, the proof shows how higher order
terms may be systematically found.

Many expressions may be found for the $_2F_1$ functions in (1.6) and (2.20) below, and
we present a sample of these in an Appendix.

A simple property of $\Delta$ is given in the following.
{\newline \bf Proposition 4}.  We have (a)
$$\int_0^1 \Delta(x)dx=-\gamma + \sum_{k=2}^\infty {{(-1)^k} \over k^2}\zeta(k),
\eqno(1.8a)$$
and (b)
$$\int_0^1 \Delta^2(x)dx=\gamma^2 -2\gamma \sum_{k=2}^\infty {{(-1)^k} \over k^2}\zeta(k)
+\sum_{m=4}^\infty {{(-1)^m} \over {(m-1)}}\sum_{\ell=2}^{m-2} {{\zeta(m-\ell)\zeta(\ell)}
\over {(m-\ell)\ell}}.  \eqno(1.8b)$$

Throughout we let $\psi^{(j)}$ denote the polygamma functions (e.g., \cite{nbs}), and we 
note the relation for integers $n >0$
$$\psi^{(n)}(x)=(-1)^{n+1}n!\zeta(n+1,x).  \eqno(1.9)$$
Therefore, as to be expected, (1.2) could equally well be written as an integral
over $\psi^{(n+1)}(xu+1)$.  The polygamma functions possess the functional equation
$$\psi^{(j)}(x+1)=\psi^{(j)}(x) + (-1)^j {{j!} \over x^{j+1}}.  \eqno(1.10)$$
For a very recent development of single- and double-integral
and series representations for the Gamma, digamma, and polygamma functions, \cite{coffeylngamma} may be consulted.

\medskip
\centerline{\bf Proof of Propositions}

{\it Proposition 1}.  We provide two alternative proofs of this result.  The result
holds for $n=0$, and for the first proof we proceed by induction.  For the inductive
step we have
$$\Delta^{(n+2)}(x)={d \over {dx}}\Delta^{(n+1)}(x)$$
$$=(-1)^n (n+1)!\int_0^1 u^{n+1} {d \over {dx}}\zeta(n+2,xu+1)du$$
$$=(-1)^{n+1} (n+2)! \int_0^1 u^{n+2} \zeta(n+3,xu+1)du.  \eqno(2.1)$$
In the last step, we used $\partial_a \zeta(s,a)=-s\zeta(s+1,a)$.  

We remark that this first method shows that (1.2) may be evaluated by repeated
integration by parts, for we have
$$(n+1)!\int_0^1 u^{n+1}\zeta(n+2,xu+1)du={{(-1)^n} \over x^n} \int_0^1 u^{n+1} \left( {\partial \over {\partial u}}\right)^n \zeta(2,xu+1)du.  \eqno(2.2)$$

Second method.  By the product rule we have
$$\Delta^{(n+1)}(x)=\sum_{j=0}^{n+1} {n \choose j} [\ln \Gamma(x+1)]^{(n-j)} {{(-1)^j j!} \over x^{j+1}}$$
$$=\sum_{j=0}^{n+1} {{n+1} \choose j} \psi^{(n-j)}(x+1) {{(-1)^j j!} \over
x^{j+1}}.  \eqno(2.3)$$
Here, it is understood that $\psi^{(-1)}(x)=\ln \Gamma(x)$.  We now apply (1.9) and the
integral representation 
$$(n-j)!\zeta(n-j+1,x+1)=\int_0^\infty {{t^{n-j} e^{-xt}} \over {e^t-1}}dt, \eqno(2.4)$$
so that
$$\Delta^{(n+1)}(x)=(-1)^{n+1}\sum_{j=0}^{n+1} {{j!} \over x^{j+1}} \int_0^\infty {{t^{n-j}
e^{-xt}} \over {e^t-1}}dt$$
$$={{(-1)^{n+1}} \over x^{n+2}} \int_0^\infty {e^{-xt} \over {(e^t-1)}}[e^{xt}\Gamma(n+2,xt)
-(n+1)!]{{dt} \over t}, \eqno(2.5)$$
where the incomplete Gamma function $\Gamma(x,y)$ has the property \cite{grad} (p. 941)
$$\Gamma(n+1,x)=n!e^{-x}\sum_{m=0}^n {x^m \over {m!}}.  \eqno(2.6)$$
Now we use a Laplace transform,
$$\int_0^1 u^{n+1}e^{-xut}du={1 \over {(xt)^{n+2}}}[(n+1)!-\Gamma(n+2,xt)], \eqno(2.7)$$
to write
$$\Delta^{(n+1)}(x)=(-1)^n \int_0^\infty {e^{-xt} \over {(e^t-1)}} t^{n+1} \int_0^1 u^{n+1}
e^{-xtu}du dt$$
$$=(-1)^n \int_0^1 u^{n+1} \int_0^\infty {t^{n+1} \over {e^t-1}} e^{-xut}dtdu$$
$$=-(-1)^n \int_0^1  u^{n+1} \zeta(n+2,xu+1)du.  \eqno(2.8)$$
By absolute convergence and the Tonelli-Hobson theorem, the interchange of integrations is
justified.  In the last step, we applied the representation (2.4).

{\it Corollary 1}.  This is proved by integrating by parts in (1.2).

{\it Remark}.  It is possible to find explicit expressions for the values $\Delta^{(n+1)}
(j+1/2)$ with half-integer argument.  This is due to the functional equation (1.10) along
with the values $\psi^{(-1)}(1/2)=\ln \sqrt{\pi}$, $\psi(1/2)=-\gamma-2\ln 2$, and
\cite{nbs} (p. 260)
$$\psi^{(n)}\left({1 \over 2}\right)=(-1)^{n+1} n!(2^{n+1}-1)\zeta(n+1), ~~~~n \geq 1.
\eqno(2.9)$$
We then obtain, for instance, by using (2.3) 
$${{\Delta^{(n+1)}\left(-{1 \over 2}\right)} \over {(n+1)!}}=\sum_{j=0}^{n-1} {{(-1)^{n-j}}\over {(n-j+1)}}(2^{n+2}-2^{j+1})\zeta(n-j+1)+2^{n+1}(\gamma+2\ln 2)-2^{n+2}\ln \sqrt{\pi}.  \eqno(2.10)$$
Similarly, it is possible to find explicit expressions for the values $\Delta^{(n+1)}
(j+1/4)$ and $\Delta^{(n+1)}(j+3/4)$ by using the corresponding values of $\psi^{(k)}$
\cite{kolbig}.

{\it Proposition 2}.  We use two Lemmas.
{\newline \bf Lemma 1}.  When the integrals involved are convergent, we have for
integrable functions $f$ and $g$
$$\int_1^\infty f(\{x\})g(x)dx=\int_0^1 f(y) \sum_{\ell=1}^\infty g(y+\ell)dy.  \eqno(2.11)$$
{\newline \bf Lemma 2}.  For $b> 0$, $\lambda>1$, and $c \geq 0$ we have for integrable
functions $f$
$$\int_0^\infty f\left(\left\{{x \over b}\right\}\right) {{dx} \over {(x+c)^\lambda}}={1 \over b^{\lambda-1}} \int_0^1 f(y)\zeta(\lambda,y+c/b)dy.  \eqno(2.12)$$
This holds when the integrals are convergent.

{\it Proof}.  For Lemma 1 we have
$$\int_1^\infty f(\{x\})g(x)dx=\sum_{\ell=1}^\infty \int_\ell^{\ell+1}f(\{x\})g(x)dx$$
$$=\sum_{\ell=1}^\infty \int_\ell^{\ell+1} f(x-\ell)g(x)dx=\sum_{\ell=1}^\infty \int_0^1 f(y)g(y+\ell) dy.  \eqno(2.13)$$
For Lemma 2 we first have 
$$\int_0^\infty f\left(\left\{{x \over b}\right\}\right) g(x)dx=b\int_0^\infty f(\{v\})g(bv)dv$$
$$=b \sum_{\ell=0}^\infty \int_\ell^{\ell+1} f(v-\ell)g(bv)dv$$
$$=b \sum_{\ell=0}^\infty \int_0^1 f(y) g[b(y+\ell)]dy.  \eqno(2.14)$$
We now put $g(x)=1/(x+c)^\lambda$, so that
$$\sum_{\ell=0}^\infty g[b(y+\ell)]={1 \over b^\lambda}\sum_{\ell=0}^\infty {1 \over {(y+\ell
+c/b)^\lambda}}={1 \over b^\lambda}\zeta(\lambda,y+c/b).  \eqno(2.15)$$

{\it Proof of Proposition 2}.  We have for integers $k \geq 1$
$$\int_0^1 u^{n+1}\zeta(n+2,ku+1)du={1 \over k^{n+2}}\int_0^k v^{n+1} \zeta(n+2,v+1)dv$$
$$={1 \over k^{n+2}}\sum_{\ell=0}^{k-1} \int_\ell^{\ell+1} v^{n+1} \zeta(n+2,v+1)dv$$
$$={1 \over k^{n+2}}\sum_{\ell=0}^{k-1} \int_0^1 (w+\ell)^{n+1}\zeta(n+2,w+\ell+1)dv.
\eqno(2.16)$$
We now apply Lemma 2 with $b=1$, $c=\ell+1$, and $f(w)=(w+\ell)^{n+1}$, giving
$$\int_0^1 u^{n+1}\zeta(n+2,ku+1)du={1 \over k^{n+2}}\sum_{\ell=0}^{k-1} \int_0^\infty
{{(\{x\}+\ell)^{n+1}} \over {(x+\ell+1)^{n+2}}}dx$$
$$={1 \over k^{n+2}}\left[\int_1^\infty {{\{w\}^{n+1}} \over w^{n+2}}dw+\sum_{\ell=1}^{k-1} \int_0^\infty {{(\{x\}+\ell)^{n+1}} \over {(x+\ell+1)^{n+2}}}dx\right].  \eqno(2.17)$$
In the last step we used the periodicity $\{w-1\}=\{w\}$.

For Corollary 2, we apply Lemma 2 of \cite{coffeylngamma}.

{\it Proposition 3}.  We start from the integral representation
$$\zeta(s,a)={a^{-s} \over 2}+{a^{1-s} \over {s-1}}-s\int_0^\infty {{P_1(x)} \over
{(x+a)^{s+1}}} dx, ~~~~\mbox{Re} ~s >-1. \eqno(2.18)$$
Then
$$\int_0^1 u^{n+1}\zeta(n+2,xu+1)du =\int_0^1 u^{n+1}\left[{1 \over {2(xu+1)^{n+2}}}+ {1 \over
{(n+1)(xu+1)^{n+1}}}\right.$$
$$\left. -(n+2)\int_0^\infty {{P_1(t)dt} \over {(t+xu+1)^{n+3}}}\right]du.  \eqno(2.19)$$
By using a standard integral representation of $_2F_1$ (e.g., \cite{grad}, p. 1040 or
\cite{andrews} p. 65) we have
$$\int_0^1 u^{n+1}\zeta(n+2,xu+1)du ={1 \over {2(n+2)}} ~_2F_1(n+2,n+2;n+3;-x)$$
$$+{1 \over {(n+1)(n+2)}} ~_2F_1(n+1,n+2;n+3;-x)
-\int_0^\infty {1 \over {(t+1)}}{{P_1(t)} \over {(t+x+1)^{n+2}}}dt.  \eqno(2.20)$$
By applying a standard transformation rule \cite{grad} (p. 1043) to the $_2F_1$ functions,
we obtain the Proposition.

{\it Corollary 3}.  We give the detailed asymptotic forms as $x \to \infty$ of the hypergeometric functions in (1.6).  We easily have that $_2F_1(1,n+1;n+3;1)=n+2$ and these forms will then show that the corresponding term in (1.6) gives the leading term as $x \to \infty$.  We let $(a)_j=\Gamma(a+j)/\Gamma(a)$ be the Pochhammer symbol.  The following expansions are valid for $|z-1|<1$ and $|\mbox{arg} (1-z)|<\pi$:
$$_2F_1(1,y;1+y;z)=y\sum_{k=0}^\infty {{(y)_k} \over {k!}}[\psi(k+1)-\psi(k+y)-\ln(1-z)]
(1-z)^k, \eqno(2.21a)$$
and
$$_2F_1(1,y;2+y;z)=y+1-y(y+1)\sum_{k=0}^\infty {{(y+1)_k} \over {k!}} [\psi(k+1)-\psi(k+y+1)-\ln(1-z)](1-z)^{k+1}, \eqno(2.21b)$$
where $(y)_0=1$.  These expansions are the $n=0$ and $n=1$ cases of (9.7.5) in \cite{lebedev}
(p. 257), respectively.  We put $y=n+1$, $z=x/(x+1)$, $\ln(1-z)=-\ln(x+1)$ and then find
$$_2F_1\left(1,n+1;n+3;{x \over {x+1}}\right)=n+2+(n+1)(n+2)[\gamma-\ln x+\psi(n+2)]{1 \over x}+O\left({{\ln x} \over x^2}\right), \eqno(2.22a)$$
and
$$_2F_1\left(1,n+2;n+3;{x \over {x+1}}\right)=-(n+2)[\gamma-\ln x+\psi(n+2)]+O\left({{\ln x}
\over x}\right). \eqno(2.22b)$$
The integral term in (1.6) is at most $O[(x+1)^{-(n+2)}]$, and is actually much smaller due
to cancellation within the integrand, and the Corollary then follows. 

{\it Remarks}.  Of course we have from (1.2) $\Delta^{(n+1)}(0)=(-1)^n (n+1)!\zeta(n+2) /(n+2)$, in agreement with the expansion (2.26).  This special case is recovered from Proposition 3 in the following way.  We have $_2F_1(a,b;c;0)=1$ and the representation \cite{titch} (p. 14)
$$\zeta(s)={1 \over {s-1}}+{1 \over 2}-s\int_1^\infty {{P_1(x)} \over x^{s+1}}dx,  \eqno(2.23)$$  
the $a=1$ case of (2.18), giving the identity $\int_0^1 u^{n+1} \zeta(n+2,1)du =\zeta(n+2)/(n+2)$.

In connection with Propositions 2 and 3, another representation that might be employed is
\cite{edwards}
$$\ln\Gamma(x+1)=\left(x+{1 \over 2}\right)\ln x-x+{1 \over 2}\ln 2\pi-\int_0^\infty {{P_1(t)}
\over {t+x}}dt.  \eqno(2.24)$$

We may note that representations (2.19) or (2.20), for instance, provide another
basis for proving integral representations for $(-1)^n\Delta^{(n+1)}(x)$ by induction.
When using (2.20), we use the derivative property
$${d \over {dx}}~_2F_1(a,b;c;-x)=-{{ab} \over c}~_2F_1(a+1,b+1;c+1;-x).  \eqno(2.25)$$
The $_2F_1$ function in (1.6) can be written in other ways, including using a
transformation formula \cite{grad} (p. 1043), so that
$$~_2F_1\left(1,n+2;n+3;{x \over {x+1}}\right)=(x+1) ~_2F_1(1,2;n+3;-x).  \eqno(2.26)$$

{\it Proposition 4}.  The result uses the expansion \cite{grad} (p. 939)
$$\ln \Gamma(x+1)=-\gamma x+\sum_{k=2}^\infty {{(-1)^k} \over k}\zeta(k) x^k.  \eqno(2.27)$$

{\it Remark}.  Let Ei$(x)$ be the exponential integral function (e.g., \cite{grad}, p. 925).
Given the relations \cite{grad} (pp. 927, 942)
$$\Gamma(0,x)=-\mbox{Ei}(-x)=-\left(\gamma+\ln x+\sum_{k=1}^\infty {{(-x)^k} \over {k k!}}
\right), \eqno(2.28)$$
it is possible to write
$$\int_0^1 \Delta(x)dx=-\gamma -\int_0^\infty {{[\gamma-t+\Gamma(0,t)+\ln t]} \over {t(e^t-1)
}}dt.  \eqno(2.29)$$
This follows by inserting a standard integral representation for the values $\zeta(k)$ into
the right side of (1.8a).  

\bigskip
\centerline{\bf Acknowledgement}
\medskip

I thank J. A. Adell for reading the manuscript.

\pagebreak
\centerline{\bf Appendix}

Here we present illustrative relations for the sort of hypergeometric functions 
appearing in (1.6) and (2.20).

The contiguous relations \cite{grad} (pp. 1044-45) may be readily applied.  As well, we
have for instance \cite{grad} (p. 1043)
$$_2F_1(n+2,n+2;n+3;-x)=(1+x)^{-(n+1)} ~_2F_1(1,1;n+3;-x).  \eqno(A.1)$$

The next result provides a type of recurrence relation in the first parameter of the
$_2F_1$ function.  
{\newline \bf Proposition A1}.  For integers $n \geq -1$ we have
$$\int_0^1 {u^{n+1} \over {(xu+1)^{n+2}}}du={1 \over {(n+2)}} ~_2F_1(n+2,n+2;n+3;-x)$$
$$={1 \over {(n+1)}}\left[{1 \over {(x+1)^{n+1}}}-{1 \over {(n+2)}}~_2F_1(n+1,n+2;n+3;-x)
\right].  \eqno(A.2)$$

{\it Proof}.  With $v=xu$ in (A.2), we have
$${1 \over x^{n+2}}\int_0^x {v^{n+1} \over {(v+1)^{n+2}}}dv={1 \over x^{n+2}}\int_0^x 
[(v+1)-v]{v^{n+1} \over {(v+1)^{n+2}}}dv$$
$$={1 \over x^{n+2}}\left[\int_0^x {v^{n+1} \over {(v+1)^{n+1}}}dv-\int_0^x {v^{n+2} \over {(v+1)^{n+2}}}dv\right]$$
$$={1 \over x^{n+2}}\left[\int_0^x {v^{n+1} \over {(v+1)^{n+1}}}dv-{{(n+2)} \over {(n+1)}}
\int_0^x {v^{n+1} \over {(v+1)^{n+1}}}dv+{1 \over {(n+1)}}{x^{n+2} \over {(x+1)^{n+1}}}
\right], \eqno(A.3)$$
where we integrated by parts.  Using a standard integral representation for $_2F_1$
\cite{grad} (p. 1040) leads to the Proposition.

Proposition A1 may be iterated in the first parameter of the $_2F_1$ function.  Then
the following relation may be applied:
$$\int_0^1 {u^{n+1} \over {(xu+1)^2}}du={1 \over {1+x}}-\left({{n+1} \over {n+2}}\right)
~_2F_1(1,n+2;n+3;-x).  \eqno(A.4)$$

The $_2F_1$ functions of concern here may be written with one or more terms containing
$\ln(x+1)$.  One way to see this is the following.  We have for the function of (A.4),
first integrating by parts,
$$\int_0^1 {u^{n+1} \over {(xu+1)^2}}du=-{1 \over x}\left[(n+1)\int_0^1 {u^n \over {xu+1}}
du+{1 \over {x+1}}\right]$$
$$=-{1 \over x}\left[{{(n+1)} \over x^{n+1}}\int_0^x {v^n \over {v+1}}dv +{1 \over {x+1}}\right]$$
$$=-{1 \over x}\left[{{(n+1)} \over x^{n+1}}\int_0^x {{[1-(1-v^n)]} \over {v+1}}dv +{1 \over {x+1}}\right]$$
$$=-{1 \over x}\left\{{{(n+1)} \over x^{n+1}}\left[\ln(x+1)-\int_0^x {{(1-v^n)} \over {v+1}}
dv\right]+{1 \over {x+1}}\right\}.  \eqno(A.5)$$
For $0\leq x \leq 1$ we may note the following simple inequality for the integral of (A.5):
$$\int_0^x {{(1-v^n)} \over {v+1}}dv \leq \int_0^x (1-v^n)dv\leq x.  \eqno(A.6)$$

\pagebreak


\begin{thebibliography}{99}
\bibitem{nbs}M. Abramowitz and I. A. Stegun,
{Handbook of Mathematical Functions, Washington, National Bureau of Standards
(1964).}
\bibitem{aa}J. A. Adell and H. Alzer,
{A monotonicity property of Euler's gamma function, Publ. Math. Debrecen (2010).}
\bibitem{andrews}G. E. Andrews, R. Askey, and R. Roy, 
{Special Functions, Cambridge University Press (1999).}
\bibitem{berg}C. Berg and H. L. Pedersen,
{Pick functions related to the gamma function, Rocky Mtn J. Math. {\bf 32}, 507-525 (2002).}
\bibitem{bochner}S. Bochner,
{Harmonic analysis and the theory of probability, Univ. California Press (1955).}
\bibitem{coffeylngamma}M. W. Coffey,
{Integral and series representations of the digamma and polygamma functions, 
arXiv:1008.0040v2 (2010).}
\bibitem{edwards}H. M. Edwards,
{Riemann's Zeta Function, Academic Press, New York (1974).}
\bibitem{grabner}P. J. Grabner, R. F. Tichy, and U. T. Zimmerman,
{Inequalities for the gamma function with applications to permanents, Discrete Math.
{\bf 154}, 53-62 (1996).}
\bibitem{grad}I. S. Gradshteyn and I. M. Ryzhik,
{Table of Integrals, Series, and Products, Academic Press, New York (1980).}
\bibitem{ivicbk}A. Ivi\'{c}, 
{The Riemann Zeta-Function, Wiley New York (1985).}
\bibitem{kolbig}K. S. K\"{o}lbig,
{The polygamma function $\psi^{(k)}(x)$ for $x={1 \over 4}$ and $x={3 \over 4}$,
J. Comput. Appl. Math. {\bf 75}, 43-46 (1996).}
\bibitem{lebedev}N. N. Lebedev,
{Special functions and their applications, Dover Publications (1972).}
\bibitem{qi}F. Qi and C.-P. Chen,
{A complete monotonicity property of the gamma function, J. Math. Analysis Appl. 
{\bf 296}, 603-607 (2004).}
\bibitem{qiguo}F. Qi and B.-N. Guo,
{Some logarithmically completely monotonic functions related to the gamma function,
{\bf 47}, 1283-1297 (2010).}
\bibitem{riemann}B. Riemann,
{\"{U}ber die Anzahl der Primzahlen unter einer gegebenen Gr\"{o}sse, 
Monats. Preuss. Akad. Wiss., 671 (1859-1860).}
\bibitem{titch}E. C. Titchmarsh,
{The Theory of the Riemann Zeta-Function, 2nd ed., Oxford University
Press, Oxford (1986).}
\bibitem{vogt}H. Vogt and J. Voigt,
{A monotonicity property of the $\Gamma$-function, J. Inequal. Pure Appl. Math.
{\bf 3}, Art. 73 (2002).}
\end{thebibliography}
\end{document}